\documentclass{ws-ijmpe}%
\usepackage{amsmath}
\usepackage{amsfonts}
\usepackage{amssymb}
\usepackage{graphicx}%
\providecommand{\U}[1]{\protect\rule{.1in}{.1in}}
\newcommand{\beq}{\vspace{0.5em}\begin{equation}}
\newcommand{\eeq}{\end{equation}\vspace{0.5em}}
\newcommand{\beqn}{\vspace{0.5em}\begin{eqnarray}}
\newcommand{\eeqn}{\end{eqnarray}\par\vspace{0.5em}\noindent}
\newcommand{\bsub}{\begin{subequations}}
\newcommand{\esub}{\end{subequations}}

\begin{document}

\title{MODERN APPLICATIONS OF COVARIANT DENSITY FUNCTIONAL
THEORY\thanks{ring@ph.tum.de}}
\author{P. Ring$^1$, H. Abusara$^2$, A. V. Afanasjev$^2$, G. A. Lalazissis$^3$,
T. Nik\v si\'c$^4$, and D. Vretenar$^4$}
\address{
$^1$ Physik-Department der Technischen Universit\"at M\"unchen, D-85748 Garching, Germany\\%
$^2$ Department of Physics and Astronomy, Mississippi State University, MS 39762\\%
$^3$ Department of Theoretical Physics, Aristotle University Thessaloniki, GR-54124, Greece\\%
$^4$ Physics Department, Faculty of Science, University of Zagreb, 10000 Zagreb, Croatia}%
\maketitle

\begin{abstract}
Modern applications of Covariant Density Functional Theory (CDFT) are discussed.
First we show a systematic investigation of fission barriers in actinide nuclei within
constraint relativistic mean field theory allowing for triaxial deformations. In
the second part we discuss a microscopic theory of quantum phase transitions (QPT)
based on the relativistic generator coordinate method.
\end{abstract}

\markboth{P. Ring et al}{Modern applications of covariant density functional theory}

\catchline{}{}{}{}{}


\section{Introduction}

Theoretical approaches based on Density Functional Theory (DFT) remain
undoubtedly among the most successful microscopic descriptions of nuclear
structure\cite{BHR.03}. One of the most successful schemes of this type is
covariant density functional theory (CDFT)\cite{Rin.96,VALR.05}. It is based
on Lorentz invariance, connecting in a consistent way the spin and spatial
degrees of freedom in the nucleus. Therefore, it needs only a relatively small
number of parameters which are adjusted to reproduce a set of bulk properties
of finite nuclei and it is valid over the entire periodic table.

Over the years, Relativistic Mean-Field (RMF) models based on the CDFT have
been successfully applied to describe ground state properties of finite
spherical and deformed nuclei over the entire nuclear chart. Excited states
have been investigated in the framework of relativistic RPA and the relativistic
Cranking model\cite{VALR.05}. Here we restrict ourselves to two more recent
applications of this concept. In section 2 we discuss a systematic investigation
of the inner fission barriers in the actinide region and in section 3 we show an
application of the relativistic Generator Coordinate Method (GCM). It allows a
microscopic description of quantum phase transitions in finite nuclei, which have
been investigated extensively in recent years by phenomenological models.

\bigskip

\section{Triaxial fission barriers in the actinide region.}

Most of the investigations of fission barriers have been carried out in the framework
of non-relativistic models\cite{SP.07}. A study of the inner fission barrier heights
$B_{f}^{st}$ of even-even nuclei is motivated by the importance of this quantity
for several physical phenomena.
Many heavy nuclei decay by spontaneous fission, and the size of the fission
barrier is a measure for the stability of a nucleus reflected in the
spontaneous fission lifetimes of these nuclei\cite{SP.07}. The probability
for the formation of a superheavy nucleus in a heavy-ion-fusion reaction is
also directly connected to the height of its fission barrier.
$B_{f}^{st}$ is a decisive quantity in the competition between neutron
evaporation and fission of a compound nucleus in the process of its cooling.
The large sensitivity of the cross section $\sigma$ for the synthesis of the
fissioning nuclei on the barrier height $B_{f}^{st}$ stresses the need for
accurate calculations of this value. For example, a change of $B_{f}^{st}$ by
1 MeV changes the calculated survival probability of a synthesized nucleus by
about one order of magnitude or even more.
The population and survival of hyperdeformed states at high spin also
depends on the fission barriers. In addition, the fission barriers of very neutron-rich nuclei have
(among other quantities such as Q-values or $\beta$-decay rates) an
influence on the $r-$process of stellar nucleosynthesis.

During the last decade the role of triaxiality in the region of the saddle
point of fission barriers has been recognized and tested in many
non-relativistic frameworks. It was found that the height of the barrier is
reduced when triaxial shapes are allowed\cite{SBD.09x,WERP.02x}.
However, this lowering strongly depends on the proton and neutron numbers and on the model
employed. Investigations of inner fission barriers with triaxiality
included are available within the frameworks of the microscopic+macroscopic
method, the extended Thomas-Fermi plus Strutinsky integral,
and non-relativistic energy density functionals based on Skyrme and Gogny
forces.


Covariant density functional theory\cite{Rin.96,VALR.05} is an approach
alternative to the above mentioned non-relativistic methods. The progress in
the study of the fission barriers within CDFT has been slower than in its
non-relativistic counterparts. Inner fission barriers in several nuclei have
been calculated mostly in the axially symmetric relativistic mean field (RMF) + BCS
approach. Many of these investigations use the constant gap approximation in the BCS part.
However, our recent study of various pairing schemes\cite{KALR.10x} used for
calculations of fission barriers clearly shows that this approximation
leads to unphysical results, because the gap changes considerably along the
fission path. Only relatively few fission barriers have been studied in 
axially symmetric RMF calculations within the BCS approximation using 
an effective density-dependent zero-range force in the pairing channel 
presenting a much more realistic approximation
for the pairing channel\cite{BRRMG.98,BBM.04x,KALR.10x}.

\begin{figure}[ptb]
\begin{center}
\includegraphics[width=8.0cm]{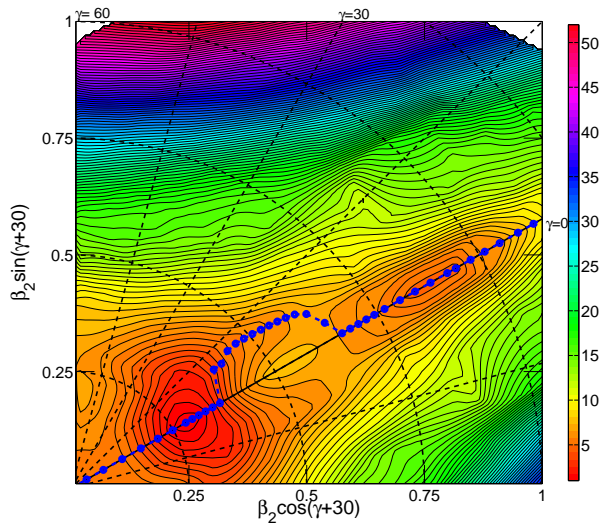}
\end{center}
\caption{(Color online) Potential energy surface in $^{240}$Pu. The energy
difference between two neighboring equipotential lines is equal to 0.5 MeV.
The blue dashed line with solid circles shows the lowest in energy solution as
a function of $\beta_{2}$. Further details are given in the text.}%
\label{fig1}
\end{figure}

Unfortunately, axially symmetric calculations cannot be directly compared with
experimental data since, as has been shown in non-relativistic
calculations\cite{SBD.09x,WERP.02x}, the lowering of fission barriers due to
triaxiality is significant and can reach 3-4 MeV in some nuclei. At present,
no systematic studies of the effects of triaxial degrees of freedom on the
height of inner fission barriers are available in covariant density functional
theory. Here we discuss recent systematic investigations\cite{AAR.10} of 
the inner fission barriers within the triaxial RMF+BCS approach, and confront 
these CDFT results with experimental quantities as far as possible.

\begin{figure}[ptb]
\begin{center}
\includegraphics[angle=0,width=12.0cm]{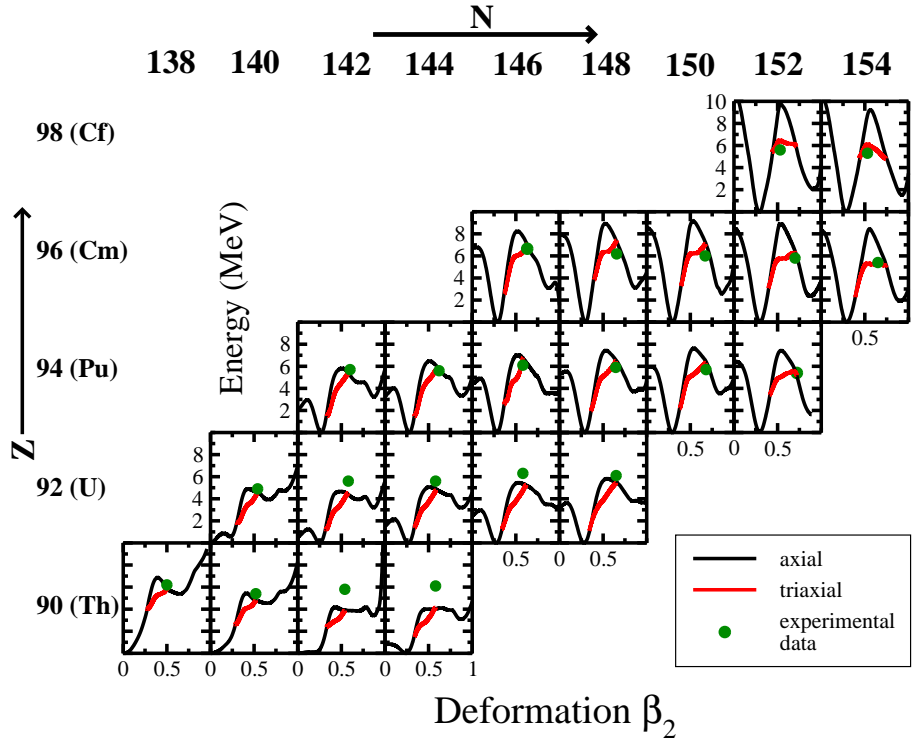}%
\end{center}%
\caption{(Color online)
Deformation energy curves of even-even actinide nuclei obtained in the RMF+BCS
calculations with the NL3* parametrization. Black solid lines display the
deformation energy curves for the axially symmetric solution, while red solid
lines show the deformation energy curves along the triaxial part of the
fission path. Solid green circles show the experimental values for the height
of inner fission barrier.}%
\label{fig2}%
\end{figure}

In this investigation we use the parameter set NL3*\cite{NL3*x} which is a
modern version of the famous set NL3 and we employ in the pairing channel a
monopole force with a set of coupling constants $G(N,Z)$  adjusted carefully
to experimental odd-even mass differences in this region\cite{AAR.10}.
In Fig.~\ref{fig1} we show as an
example the potential energy surface of the nucleus $^{240}$Pu in the
$\beta$-$\gamma$ plane. For axial symmetry we find the normal deformed minimum of
the ground state at a deformation $\beta\sim0.28$, a maximum at $\beta
\sim0.52$ and a superdeformed minimum at $\beta\sim0.96$. We observe that the
fission path (the blue dashed curve) bypasses the axial barrier between the normal and superdeformed
minima. The barrier height is determined by the maximum of the energy along
this fission path.

The deformation energy curves for other even-even nuclei in this region
obtained in these calculations are shown in Fig.~\ref{fig2}. Full black curves
show axially symmetric solutions, while the values of the deformation
energy curves along the triaxial fission path are given in red. One can see
that by allowing for triaxial deformation the fission barrier heights are
reduced by $1-4$ MeV as compared with axially symmetric solutions. This
lowering depends on the proton and neutron numbers. It also brings in average
the results of the calculations in closer agreement with experimental data
shown by green solid circles in Fig.~\ref{fig2}. These circles display the
height of the experimental fission barrier\cite{SGP.05} at the calculated $\beta
$-deformation of the saddle point. A typical uncertainty in the experimental values, as
suggested by the differences among various compilations, is of the order of
$\pm0.5$ MeV.

\begin{figure}[ptb]
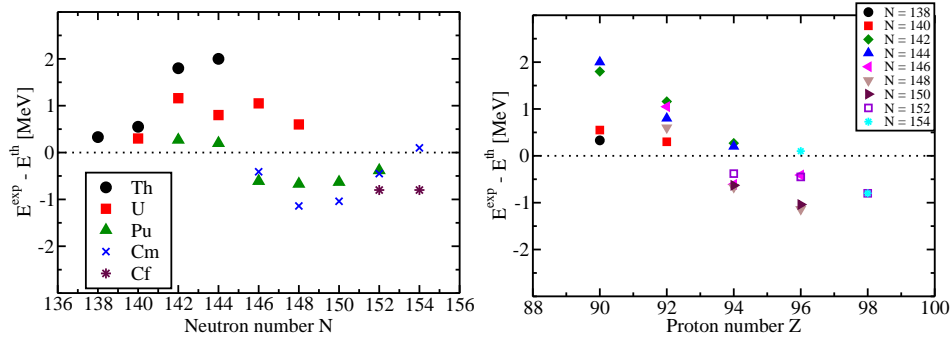

\begin{center}%
\includegraphics[width=6.2cm]{fig3a.eps}
\includegraphics[width=6.2cm]{fig3b.eps}
\end{center}%
\caption{(Color online) The difference
between experimental and calculated heights of inner fission barriers as a
function of neutron number $N$ and proton number $Z$}%
\label{fig3}%
\end{figure}

The microscopic origin of the lowering of the barrier due to triaxiality can
be traced back to the changes of the level density in the vicinity of the
Fermi level induced by triaxiality. The single-particle level density at the
Fermi level is lower for triaxial solutions than for axially symmetric
solutions. This is especially important at the deformation corresponding to
the saddle point of the axially symmetric solution. Fig.~\ref{fig3} shows
the differences between calculated and experimental heights of inner fission
barriers. The average deviation between theory and experiment is 0.76 MeV.
This is comparable with the results obtained in the macroscopic+microscopic
method.

\section{Microscopic theory of quantum phase transitions.}

Applications of CDFT discussed so far are based on the mean field concept.
However, this framework has its limitations. It is characterized by
symmetry breaking. and can only provide an approximate description of specific
nuclear excitations, such as collective rotations in the framework of the
Cranking model or vibrations in the framework of time-dependent mean field
theory, which corresponds in the small amplitude limit to RPA or QRPA. To
calculate excitation spectra with specific quantum numbers and electromagnetic
transition rates in individual nuclei, it is necessary to extend this
framework to include collective correlations related to the restoration of
broken symmetries and to fluctuations of collective coordinates. Collective
correlations are sensitive to shell effects, display pronounced variations
with particle number, and cannot be incorporated in a single Slater determinant.

On this level an extension of density functional theory beyond mean field takes
into account collective correlations through the restoration of broken symmetries
and configuration mixing of symmetry-breaking product states. In recent years several accurate
and efficient models and algorithms have been developed that perform the
restoration of symmetries broken by the static nuclear mean field, and take
into account fluctuations around the mean-field minimum. The most effective
approach to configuration mixing calculations is the generator coordinate
method (GCM)\cite{RS.80} based on the variational ansatz for the
many-body wave function:
\begin{equation}
\left\vert \Psi_{\alpha}\right\rangle =\int dq{f_{\alpha}(q)P}^{I}P^{N}%
P^{Z}{\left\vert \Phi(q)\right\rangle }\;.
\label{GCM-wave}
\end{equation}
The set of intrinsic wave functions $|\Phi(q)\rangle$ is generated by
self-consistent constraint RMF+BCS calculations. The weight functions
$f_{\alpha}(q)$ in the collective wave function
of Eq.~(\ref{GCM-wave}) are determined from the variation of the energy by the
Hill-Wheeler equation: \vspace{0.5em}%
\begin{equation}
\int dq^{\prime}\left[  \mathcal{H}(q,q^{\prime})-E_{\alpha}\mathcal{N}%
(q,q^{\prime})\right]  f_{\alpha}(q^{\prime})=0,
\end{equation}
\vspace{0.5em}where $\mathcal{H}$ and $\mathcal{N}$ are the angular-momentum
projected GCM kernel matrices of the Hamiltonian and the norm,
respectively.With the simplifying assumption of axial symmetry, GCM
configuration mixing of angular-momentum, and even particle-number projected
quadrupole-deformed mean-field states, has become a standard tool in nuclear
structure studies with non-relativistic\cite{VHB.00,RER.02b}, and
relativistic density functionals\cite{NVR.06a,NVR.06b}.

Much more involved and technically difficult is the description of intrinsic
quadrupole modes including triaxial deformations. Only very recently fully
microscopic three-dimensional calculations have been presented for
Skyrme\cite{BH.08} and Gogny\cite{RE.10} mean field models generated by
triaxial quadrupole constraints that are projected and mixed by the generator
coordinate method. This method is actually equivalent to a seven-dimensional
GCM calculation, mixing all five degrees of freedom of the quadrupole operator
and the gauge angles for protons and neutrons. Only very recently 3D angular
momentum projection\cite{YMRV.10} and configuration mixing of
angular-momentum projected triaxial relativistic mean-field wave
functions\cite{YMRV.10,YMC.10} has been possible. However, triaxial calculations of
this type are connected with an extreme numerical effort and are, at present,
not possible for heavy nuclei.

Here we report on an axially symmetric application, where the set
$|\Phi(q)\rangle$ is generated by imposing constraints on the axial $q_{20}$ mass
quadrupole moments in self-consistent RMF+BCS calculations. The following
calculations are based on the relativistic point coupling model
PC-F1\cite{BMM.02x}. Pairing correlations are taken into account in the
simple BCS approximation, using a $\delta$-pairing force.

\begin{figure}[ptb]
\vspace{-1cm}
\begin{center}
\includegraphics[width=10cm,angle=270]{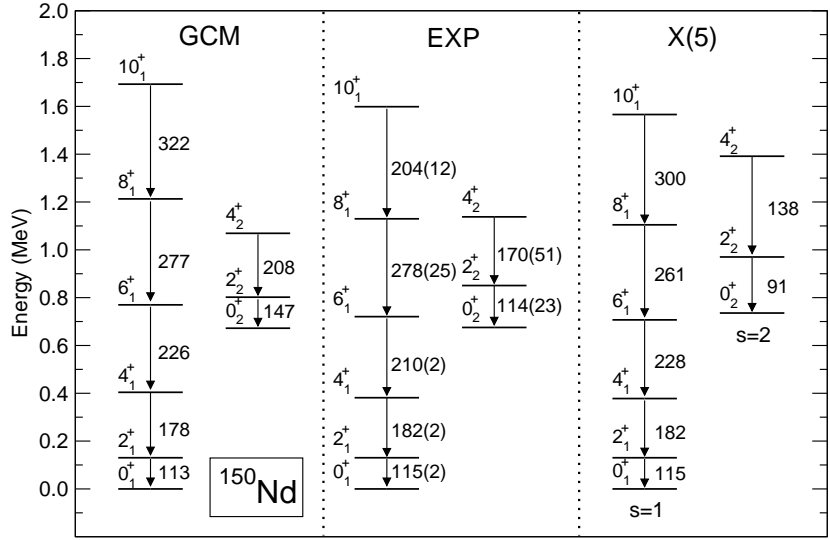}
\end{center}
\vspace{-1cm}
\caption{The particle-number projected GCM spectrum of $^{150}$Nd (left),
compared with the and the X(5)-symmetry predictions (right) for the excitation
energies, and B(E2) values (in Weisskopf units) of the ground-state and the
$\beta$-band. The theoretical spectra are normalized to the experimental
energy of the state 2$^{+}_{1}$, and the X(5) transition strengths are
normalized to the experimental B(E2;$2^{+}_{1}\rightarrow0^{+}_{1}$).}%
\label{fig4}%
\end{figure}

Quantum phase transitions from spherical to deformed shapes have recently
gained considerable interest. However, most of these investigations have
been carried out within simple boson models\cite{Iac.01}. In order to understand
the microscopic background of such investigations in finite nuclei the
relativistic Generator Coordinate Method has recently been applied\cite{NVLR.07}.
Starting from constrained self-consistent mean-field calculations of potential
energy curves, the one-dimensional generator coordinate method (GCM) has been
used to perform configuration mixing of angular-momentum and particle-number
projected relativistic wave functions\cite{NVLR.07}. It has been shown that
the microscopic framework based on universal density functionals, adjusted
only to nuclear ground-state properties and extended to take into account
correlations related to symmetry restoration and fluctuations of collective
variables, is able to describe not only general features of such shape
transitions but also the unique behavior of excitation spectra and
transition rates at the critical points. In particular the shape phase
transition in the chain of Nd isotopes occurs precisely at the isotope
$^{150}$Nd, in agreement with empirical evidence for the realization
of the X(5) symmetry in the N = 90 rare-earth isotones.
As shown in Fig.~\ref{fig4} the calculated spectrum and the intra-band
and inter-band B(E2) transitions for $^{150}$Nd are in
excellent agreement with data\cite{Kru.02} and close to the X(5)-model
predictions for the nucleus $^{150}$Nd\cite{Iac.01}.

As discussed before, because of the considerable numerical effort
3D angular momentum projected GCM calculations are at present not
yet possible for such heavy nuclei as the Nd-chain. Therefore this
phase transition has been investigated recently\cite{LNV.09a} on the
basis of a 5-dimensional Bohr Hamiltonian, whose
parameters were derived microscopically form the covariant density functional
PC-F1. The resulting spectrum and the electromagnetic transition probabilities
are more or less the same as in the 1D-AMP GCM calculation shown in
Fig.~\ref{fig4}.

\section{Summary and outlook.}

Among the microscopic approaches to the nuclear many-body problem, the
framework of nuclear density functional theory provides the most complete
description of ground-state properties and collective excitations over the
whole nuclide chart. Here we have seen that the self-consistent relativistic
mean-field model based on a universal density functional provides an excellent
description of the inner fission barriers in actinide nuclei with an average
deviation between theory and experiment of 0.76 MeV. For transitional nuclei
relativistic models not only describe general features of shape transitions
but also particular properties of spectra and transition rates at
the critical point of the QPT. However, to calculate excitation spectra and
transition probabilities, the self-consistent mean-field approach has to be
extended to include correlations related to restoration of broken symmetries
and fluctuations of collective variables. This can be done either by
performing GCM configuration mixing calculations of projected wave functions
or by constructing collective Bohr-type Hamiltonians with
deformation-dependent parameters determined from self-consistent mean-field
calculations. The possibility to perform self-consistent microscopic studies
of shape transitions opens a new perspective on the origin of nuclear quantum
phase transitions in various mass regions. It is therefore important to
systematically analyze these methods, also employing different energy-density
functionals, various types of shape phase transitions that have been predicted
in several regions of medium-heavy and heavy nuclei.

\section*{Acknowledgments}

This work was partly supported by the DFG cluster of excellence
\textquotedblleft Origin and Structure of the Universe\textquotedblright\ (www.universe-cluster.de).



\end{document}